\documentclass[11pt,twoside]{article}  
\usepackage[ansinew]{inputenc}
\usepackage{amsfonts}
\usepackage{amssymb,amsmath}
\usepackage{latexsym}

\newcommand {\debeq}	{\begin{eqnarray*}}
\newcommand {\fineq}	{\end{eqnarray*}}
\newcommand {\lbd}	{\lambda}

\newcommand	{\tendinfty}	
{\rightarrow\infty}

\newcommand	{\intgen}	
{\int_0^\infty}

\newcommand	{\PP}{\mathbb{P}}
\newcommand	{\EE}{\mathbb{E}}

\newtheorem	{thm}		{Theorem}[section]

\newtheorem	{lem} 	[thm]	{Lemma}

\newtheorem     {rem}           {Remark}
\newtheorem	{prop}	[thm]{Proposition}
\newtheorem	{cor}		[thm]{Corollary}

\begin{document}

\title{Species abundance distributions in neutral models with immigration or mutation and general lifetimes.}
\author{\textsc{By Amaury Lambert, UPMC Univ Paris 06
}
}
\date{\today}
\maketitle
\noindent\textsc{Laboratoire de Probabilités et Modèles Aléatoires\\
UMR 7599 CNRS and UPMC Univ Paris 06\\
Case courrier 188\\
4, Place Jussieu\\
F-75252 Paris Cedex 05, France}\\
\textsc{E-mail: }amaury.lambert@upmc.fr\\
\textsc{URL: }http://www.proba.jussieu.fr/pageperso/amaury/index.htm

\begin{abstract}
\noindent
We consider a general, neutral, dynamical model of biodiversity.
Individuals have i.i.d. lifetime durations, which are not necessarily exponentially distributed, and each individual gives birth independently at constant rate $\lambda$. Thus, the population size is a \emph{homogeneous, binary Crump--Mode--Jagers process} (which is not necessarily a Markov process). We assume that types are clonally inherited.

We consider two classes of speciation models in this setting. In the \emph{immigration model}, new individuals of an entirely new species singly enter  the population at constant rate $\mu$ (e.g., from the mainland into the island). In the \emph{mutation model}, each individual independently experiences point mutations in its germ line, at constant rate $\theta$.

We are interested in the \emph{species abundance distribution},  i.e., in the numbers, denoted $I_n(k)$ in the immigration model and $A_n(k)$ in the mutation model, of species represented by $k$ individuals, $k=1,2,\ldots,n$, when there are $n$ individuals in the total population.

In the immigration model, we prove that the numbers $(I_t(k);k\ge 1)$ of species represented by $k$ individuals \emph{at time $t$}, are independent Poisson variables with parameters as in {Fisher's log-series}. When conditioning on the total size of the population to equal $n$, this results in species abundance distributions given by \emph{Ewens' sampling formula}. In particular, $I_n(k)$ converges as $n\to\infty$ to a Poisson r.v. with mean $\gamma /k$, where $\gamma:=\mu/\lambda$.

In the mutation model, as $n\to\infty$, we obtain the almost sure convergence of $n^{-1}A_n(k)$ to a nonrandom explicit constant. In the case of a critical, linear birth--death process, this constant is given by Fisher's log-series,  namely $n^{-1}A_n(k)$ converges to $\alpha^{k}/k$, where $\alpha :=\lambda/(\lambda+\theta)$. 

In both models, the abundances of the most abundant species are briefly discussed.

\end{abstract}  	
\medskip
\textit{Running head.} Neutral models of biodiversity with general lifetimes.

\section{Introduction}

Our goal is to study two models of speciation  in the vein of the neutral theory of biodiversity \cite{H}, an \emph{immigration model} and a \emph{mutation model}, both in a same general birth/death dynamical setting. A specific feature of our results is that no assumption is made on the distribution of lifetime durations, contrasting with usual Markovian dynamics where this distribution is exponential.

We assume that particles behave independently from one another, that each particle gives birth at constant rate $\lbd$ during its lifetime (interbirth durations are i.i.d. exponential random variables with parameter $\lbd$), and that lifetime durations are i.i.d.. Then the process $(N_t;t\ge 0)$ giving the number of extant individuals at time $t$, belongs to a wide class of branching processes called \emph{Crump--Mode--Jagers processes}. Actually, the processes we consider are homogeneous (constant birth rate) and binary (one birth at a time) but differ in generality from classic birth--death processes in that the lifetimes durations may follow a general distribution. \\
\\
Now each individual bears some type (or, equivalently, belongs to some species), and we will assume that, at each birth time $t$, the type of the mother at time $t$ is passed on to their offspring without modification. However, new species can arise in this population. These new types can arise in two fashions, whence defining either speciation model.

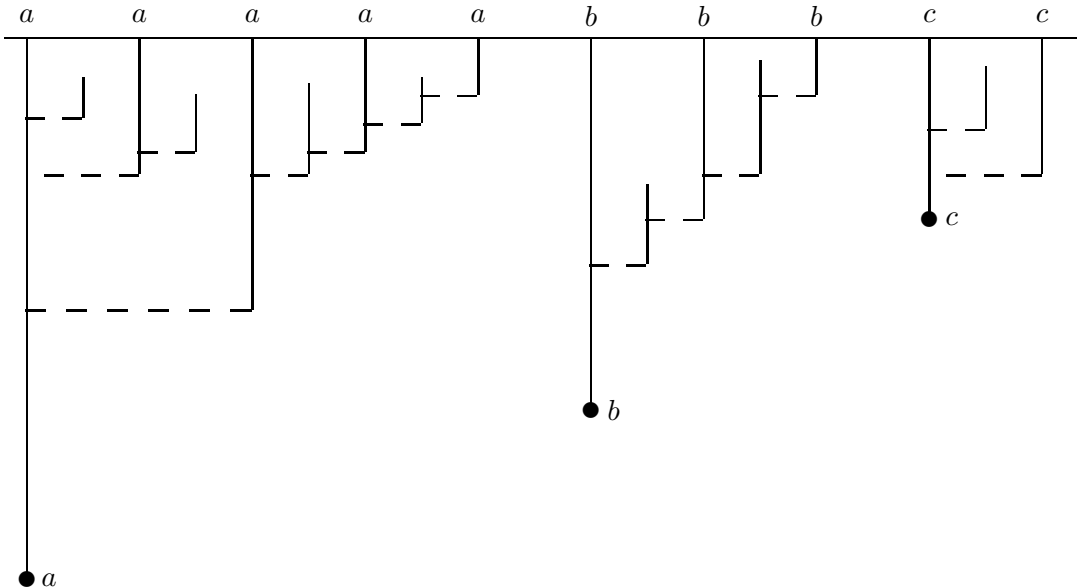
\begin{figure}[ht]
\label{fig : mig}
\unitlength 3mm 
\linethickness{0.5pt}

\begin{picture}(50.5,33)(3,-2)
\put(12,30){\line(0,-1){6}}
\put(52,30){\line(0,-1){6}}
\put(7,30){\line(0,-1){24}}
\put(17,30){\line(0,-1){12}}
\put(32,30){\line(0,-1){16.5}}
\put(7,6){\circle*{.707}}
\put(47,22){\circle*{.707}}
\put(32,13.5){\circle*{.707}}
\put(11.93,23.93){\line(-1,0){.8333}}
\put(10.263,23.93){\line(-1,0){.8333}}
\put(8.596,23.93){\line(-1,0){.8333}}
\put(51.93,23.93){\line(-1,0){.8333}}
\put(50.263,23.93){\line(-1,0){.8333}}
\put(48.596,23.93){\line(-1,0){.8333}}
\put(48,22){\makebox(0,0)[cc]{$c$}}
\put(7,31){\makebox(0,0)[cc]{$a$}}
\put(12,31){\makebox(0,0)[cc]{$a$}}
\put(17,31){\makebox(0,0)[cc]{$a$}}
\put(22,31){\makebox(0,0)[cc]{$a$}}
\put(27,31){\makebox(0,0)[cc]{$a$}}
\put(47,31){\makebox(0,0)[cc]{$c$}}
\put(52,31){\makebox(0,0)[cc]{$c$}}
\put(9.5,26.5){\line(0,1){1.75}}
\put(14.5,25){\line(0,1){2.5}}
\put(9.43,26.43){\line(-1,0){.8333}}
\put(7.763,26.43){\line(-1,0){.8333}}
\put(14.43,24.93){\line(-1,0){.8333}}
\put(12.763,24.93){\line(-1,0){.8333}}
\put(21.93,24.93){\line(-1,0){.8333}}
\put(20.263,24.93){\line(-1,0){.8333}}
\put(22,30){\line(0,-1){5}}
\put(19.5,28){\line(0,-1){4}}
\put(19.43,23.93){\line(-1,0){.8333}}
\put(17.763,23.93){\line(-1,0){.8333}}
\put(39.5,29){\line(0,-1){5}}
\put(39.43,23.93){\line(-1,0){.8333}}
\put(37.763,23.93){\line(-1,0){.8333}}
\put(41.93,27.43){\line(-1,0){.8333}}
\put(40.263,27.43){\line(-1,0){.8333}}
\put(26.93,27.43){\line(-1,0){.8333}}
\put(25.263,27.43){\line(-1,0){.8333}}
\put(27,30){\line(0,-1){2.5}}
\put(24.5,28.25){\line(0,-1){2}}
\put(24.43,26.18){\line(-1,0){.8333}}
\put(22.763,26.18){\line(-1,0){.8333}}
\put(46.93,25.93){\line(1,0){.8333}}
\put(48.596,25.93){\line(1,0){.8333}}
\put(34.5,23.5){\line(0,-1){3.5}}
\put(34.43,21.93){\line(1,0){.8333}}
\put(36.096,21.93){\line(1,0){.8333}}
\put(37,30){\line(0,-1){8}}
\put(37,31){\makebox(0,0)[cc]{$b$}}
\put(16.93,17.93){\line(-1,0){.9091}}
\put(15.112,17.93){\line(-1,0){.9091}}
\put(13.293,17.93){\line(-1,0){.9091}}
\put(11.475,17.93){\line(-1,0){.9091}}
\put(9.657,17.93){\line(-1,0){.9091}}
\put(7.839,17.93){\line(-1,0){.9091}}
\put(47,22){\line(0,1){8}}
\put(49.5,26){\line(0,1){2.75}}
\put(34.43,19.93){\line(-1,0){.8333}}
\put(32.763,19.93){\line(-1,0){.8333}}
\put(6,30){\line(1,0){47.5}}
\put(33,13.5){\makebox(0,0)[cc]{$b$}}
\put(8,6){\makebox(0,0)[cc]{$a$}}
\put(32,31){\makebox(0,0)[cc]{$b$}}
\put(42,27.5){\line(0,1){2.5}}
\put(42,31){\makebox(0,0)[cc]{$b$}}
\end{picture}

\caption{The immigration model. Time axis is vertical; horizontal axis shows filiation. Solid dots show the arrival times of immigrants, who all have distinct types labelled by letters $a,b,c$. The type of each extant individual  is also shown.}
\end{figure}

The immigration model is a generalization of Karlin and McGregor's model \cite{KMcG} to general lifetimes. It intends to model a population on an island receiving immigrants from the mainland, as in the theory of island biogeography \cite{McAW}. We assume that new propagules  singly enter the island population at the instants of a Poisson process with rate $\mu$, called the \emph{immigration rate}, and behave from then on, as the other particles on the island. Each of these immigrating particles is of an entirely new species, but their whole descendance is entirely clonal. See Figure \ref{fig : mig}.

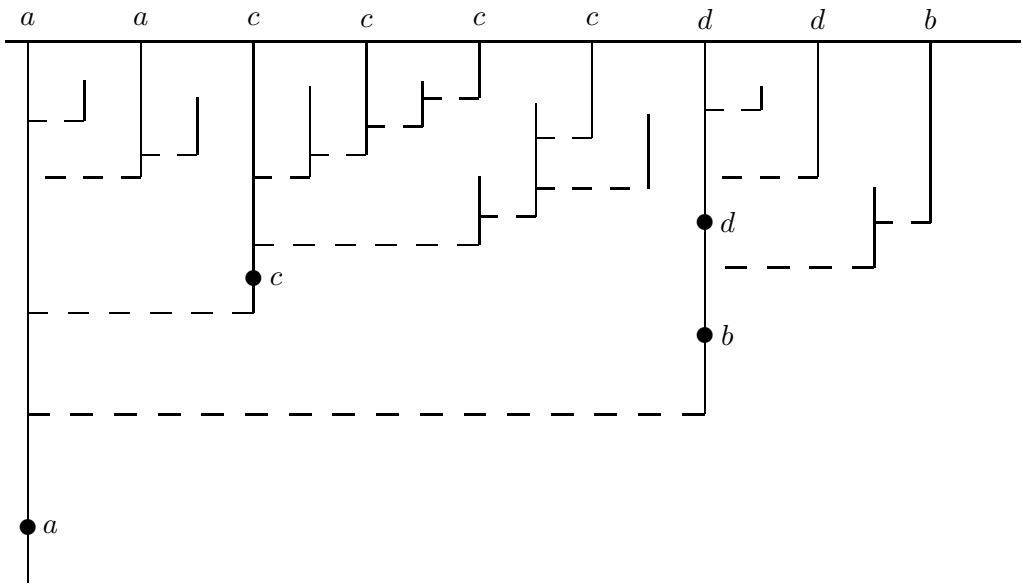
\begin{figure}[ht]
\unitlength 3mm 
\linethickness{0.5pt}
\ifx\plotpoint\undefined\newsavebox{\plotpoint}\fi 
\begin{picture}(48,33)(3,-2)
\put(6,30){\line(1,0){45}}
\put(12,30){\line(0,-1){6}}
\put(42,30){\line(0,-1){6}}
\put(7,30){\line(0,-1){24}}
\put(17,30){\line(0,-1){12}}
\put(37,30){\line(0,-1){16.5}}
\put(7,8.5){\circle*{.707}}
\put(17,19.5){\circle*{.707}}
\put(37,22){\circle*{.707}}
\put(37,17){\circle*{.707}}
\put(11.971,23.971){\line(-1,0){.8333}}
\put(10.304,23.971){\line(-1,0){.8333}}
\put(8.637,23.971){\line(-1,0){.8333}}
\put(16.971,17.971){\line(-1,0){.9091}}
\put(15.153,17.971){\line(-1,0){.9091}}
\put(13.334,17.971){\line(-1,0){.9091}}
\put(11.516,17.971){\line(-1,0){.9091}}
\put(9.698,17.971){\line(-1,0){.9091}}
\put(7.88,17.971){\line(-1,0){.9091}}
\put(36.971,13.471){\line(-1,0){.9677}}
\put(35.035,13.471){\line(-1,0){.9677}}
\put(33.1,13.471){\line(-1,0){.9677}}
\put(31.164,13.471){\line(-1,0){.9677}}
\put(29.229,13.471){\line(-1,0){.9677}}
\put(27.293,13.471){\line(-1,0){.9677}}
\put(25.358,13.471){\line(-1,0){.9677}}
\put(23.422,13.471){\line(-1,0){.9677}}
\put(21.487,13.471){\line(-1,0){.9677}}
\put(19.551,13.471){\line(-1,0){.9677}}
\put(17.616,13.471){\line(-1,0){.9677}}
\put(15.68,13.471){\line(-1,0){.9677}}
\put(13.745,13.471){\line(-1,0){.9677}}
\put(11.809,13.471){\line(-1,0){.9677}}
\put(9.874,13.471){\line(-1,0){.9677}}
\put(7.938,13.471){\line(-1,0){.9677}}
\put(41.971,23.971){\line(-1,0){.8333}}
\put(40.304,23.971){\line(-1,0){.8333}}
\put(38.637,23.971){\line(-1,0){.8333}}
\put(8,8.5){\makebox(0,0)[cc]{$a$}}
\put(18,19.5){\makebox(0,0)[cc]{$c$}}
\put(38,22){\makebox(0,0)[cc]{$d$}}
\put(38,17){\makebox(0,0)[cc]{$b$}}
\put(7,31){\makebox(0,0)[cc]{$a$}}
\put(12,31){\makebox(0,0)[cc]{$a$}}
\put(17,31){\makebox(0,0)[cc]{$c$}}
\put(27,31){\makebox(0,0)[cc]{$c$}}
\put(32,31){\makebox(0,0)[cc]{$c$}}
\put(37,31){\makebox(0,0)[cc]{$d$}}
\put(42,31){\makebox(0,0)[cc]{$d$}}
\put(9.5,26.5){\line(0,1){1.75}}
\put(14.5,25){\line(0,1){2.5}}
\put(9.471,26.471){\line(-1,0){.8333}}
\put(7.804,26.471){\line(-1,0){.8333}}
\put(14.471,24.971){\line(-1,0){.8333}}
\put(12.804,24.971){\line(-1,0){.8333}}
\put(16.971,20.971){\line(1,0){.9091}}
\put(18.789,20.971){\line(1,0){.9091}}
\put(20.607,20.971){\line(1,0){.9091}}
\put(22.425,20.971){\line(1,0){.9091}}
\put(24.243,20.971){\line(1,0){.9091}}
\put(26.062,20.971){\line(1,0){.9091}}
\put(27,21){\line(0,1){3}}
\put(21.971,24.971){\line(-1,0){.8333}}
\put(20.304,24.971){\line(-1,0){.8333}}
\put(22,30){\line(0,-1){5}}
\put(19.5,28){\line(0,-1){4}}
\put(19.471,23.971){\line(-1,0){.8333}}
\put(17.804,23.971){\line(-1,0){.8333}}
\put(32,30){\line(0,-1){4.25}}
\put(29.5,27.25){\line(0,-1){5}}
\put(29.471,22.221){\line(-1,0){.8333}}
\put(27.804,22.221){\line(-1,0){.8333}}
\put(31.971,25.721){\line(-1,0){.8333}}
\put(30.304,25.721){\line(-1,0){.8333}}
\put(26.971,27.471){\line(-1,0){.8333}}
\put(25.304,27.471){\line(-1,0){.8333}}
\put(27,30){\line(0,-1){2.5}}
\put(24.5,28.25){\line(0,-1){2}}
\put(24.471,26.221){\line(-1,0){.8333}}
\put(22.804,26.221){\line(-1,0){.8333}}
\put(29.471,23.471){\line(1,0){.8333}}
\put(31.137,23.471){\line(1,0){.8333}}
\put(32.804,23.471){\line(1,0){.8333}}
\put(34.5,23.5){\line(0,1){3.25}}
\put(36.971,26.971){\line(1,0){.8333}}
\put(38.637,26.971){\line(1,0){.8333}}
\put(39.5,27){\line(0,1){1}}
\put(44.5,23.5){\line(0,-1){3.5}}
\put(44.471,19.971){\line(-1,0){.9375}}
\put(42.596,19.971){\line(-1,0){.9375}}
\put(40.721,19.971){\line(-1,0){.9375}}
\put(38.846,19.971){\line(-1,0){.9375}}
\put(44.471,21.971){\line(1,0){.8333}}
\put(46.137,21.971){\line(1,0){.8333}}
\put(47,30){\line(0,-1){8}}
\put(47,31){\makebox(0,0)[cc]{$b$}}
\put(22,30.938){\makebox(0,0)[cc]{$c$}}
\end{picture}

\caption{The mutation model. Time axis is vertical; horizontal axis shows filiation. Solid dots show the mutation events. Each mutation yields a new type, labelled by letters $a,b,c,d$. The type of each extant individual  is also shown.}
\label{fig : mut}

\end{figure}

In the mutation model, we assume that the germ line of each particle experiences mutations during the whole lifetime of the particle. At the instants of a Poisson process with rate $\theta$, the type of the particle changes to an entirely new type, as in the \emph{infinitely-many alleles model} \cite{Ewens}. See Figure \ref{fig : mut}.

Another way of seeing the model is to replace the word particle with the word colony, and the word population with the word metapopulation. Then in our model, all individuals of a colony are of the same species, lifetimes are extinction times of colonies, and birth events correspond to propagules sent out by a colony to found a brand new colony. Immigration events correspond to propagules immigrating from the mainland and founding simultaneously a brand new colony. Mutation events correspond to mutants appearing in a colony and getting to fixation instantaneously. This way of modeling speciation is more satisfactory, but we stick to the first terminology not to obscure reading.

\section{Statements of results and Fisher's logarithmic series}

In \cite{F, FCW}, R.A. Fisher and his coauthors suggested a simple model of species count where the probability of observing $k$ individuals  of a given species is $c\alpha^k/k$ for some constant $\alpha\in(0,1)$. Following this, a number of authors proposed dynamical models where this so-called \emph{log-series} not only gives the distribution of the number of individuals of a single species, but also the multivariate species abundance distribution of a community, in the sense that the number of species represented by $k$ individuals follows independently a Poisson distribution with parameter  $c\alpha^k/k$. For example, Karlin and McGregor \cite{KMcG} studied various dynamical models of structured populations, including a critical birth--death process with immigration which is a particular case of our immigration model (i.e., where the lifespan is exponentially distributed), satisfying the previously described property. See also \cite{Ken, KC}, and \cite{W} for a very nice and comprehensive account on these models and on their associated multivariate distributions.\\

Let us fix some time $t$. In the immigration model (resp. in the mutation model), we let $I_t(k)$ (resp. $A_t(k)$) denote the number of species represented by $k$ individuals at time $t$. When conditioning on the total number of individuals being $n$ at this fixed time $t$, we will write $I_t(k)$ instead of $I_n(k)$ and $A_t(k)$ instead of $A_n(k)$.
The vectors $(I_.(k))_k$ and $(A_.(k))_k$ are called \emph{frequency spectra}.

In the immigration model, we actually provide a rather accurate result (Theorem \ref{thm : spectrum mig}) on the spectrum at any time $t$, without conditioning on the number of individuals, stating that the random variables $(I_t(k))_k$ are independent Poisson variables with parameters as in Fisher's log-series, with a parameter $\alpha$ depending on time $t$.  In Corollary \ref{cor : Poisson cond}, we prove that the random vector $(I_n(1), \ldots, I_n(n))$ has the same law as a vector of independent Poisson variables $(Y_1,\ldots, Y_n)$ conditioned on $\sum_{k=1}^n kY_k = n$, where $Y_k$ follows the Poisson distribution with parameter $\gamma/ k$, $\gamma$ being defined as the immigration-to-birth rate ratio $\mu/\lbd$. 
These two results are known in the case of a critical, linear birth--death process \cite{KMcG}. Notice that the conditioning in the corollary not only removes the dependence upon the origination time $t$, but also on the distribution of lifetime durations.
This spectrum is exactly the one described by \emph{Ewens' sampling formula} \cite{D,Ew, Ewens}. The asymptotic behaviour of this spectrum is well-known (see for example \cite{DT, D}): for any fixed $j$,
$$
\lim_{n\to\infty} (I_n(1), I_n(2),\ldots,I_n(j))\stackrel{\cal L}{=} (Y_1,Y_2, \ldots, Y_j)
$$
where the $Y_k$'s are \emph{independent} Poisson variables with parameter $\gamma/ k$. 

This result contrasts with the mutation model, where species with abundance $k$ are shown to accumulate linearly with population size, instead of stabilizing as previously. First, Theorem \ref{thm : exp freq spectrum} gives the expected number of species with a fixed age and with abundance $k$. Then Theorem \ref{thm : spectrum mut} gives exact formulae for the almost-sure asymptotic accumulation of species with given abundances. In the case of a critical birth--death process with (birth/death rate $\lbd$ and) mutation rate $\theta$, we get 
$$
\lim_{n\tendinfty} n^{-1}A_n(k) 	=	c\frac{\alpha^k}{k} \qquad \mbox{a.s.},
$$
where $\alpha:=\lbd/(\lbd +\theta)$, and $c=(1-\alpha)/\alpha$. We also have the a.s. convergence of the total number of species $A_n$ divided by $n$ to $-c\ln (1-\alpha)$. 

Thus, species with $k$ individuals tend to accumulate linearly with sample size in the mutation model, while their cardinality converges to a finite random variable in the immigration model. This has an important consequence for the species with a large number of individuals. In the immigration model, it can be shown that the  oldest $j$ species on the island have a number of individuals of the order of $n$, as $n$ grows \cite{Richard}. In the mutation model, in contrast, the proportion $B_n(k)$ of individuals belonging to species with more than $k$ individuals is
$$
B_n(k)=1-n^{-1}\sum_{j=1}^{k-1}jA_n(j)\longrightarrow 1-\sum_{j=1}^{k-1}c\alpha^j=1-(1-\alpha)\sum_{j=1}^{k-1}\alpha^{j-1}=\alpha^{k-1}.
$$
As a consequence, for any $\varepsilon >0$, there is an integer $k$ such that $\limsup_n B_n(k)\le\varepsilon$. Actually, independent calculations \cite{CL2} show that the most abundant species have abundances of the order of $n^\beta$, with $\beta=1-\theta/\eta$, where $\eta$ is the exponential growth rate of the total population, 
 in the case when the mutation rate $\theta$ is smaller than $\eta$. In the case when $\theta>\eta$, these abundances are of the order of $\log(n)$.

\section{Splitting trees and coalescent point processes}

The genealogical trees  that we consider here are usually called splitting trees \cite{GK}. Splitting trees are those random trees where individuals give birth at constant rate $\lambda$ during a lifetime with general distribution $\pi(\cdot)/\lambda$, to i.i.d. copies of themselves, where $\pi$ is a positive measure on $(0,\infty]$ with total mass $\lambda$ called the \emph{lifespan measure}. We assume that they are started with one unique progenitor born at time 0. We denote by $\PP$ their law, and the subscript $s$ in $\PP_s$ means conditioning on the lifetime of the progenitor being $s$. Of course if $\PP$ bears no subscript, this means that the lifetime of the progenitor follows the usual distribution $\pi(\cdot)/\lambda$.

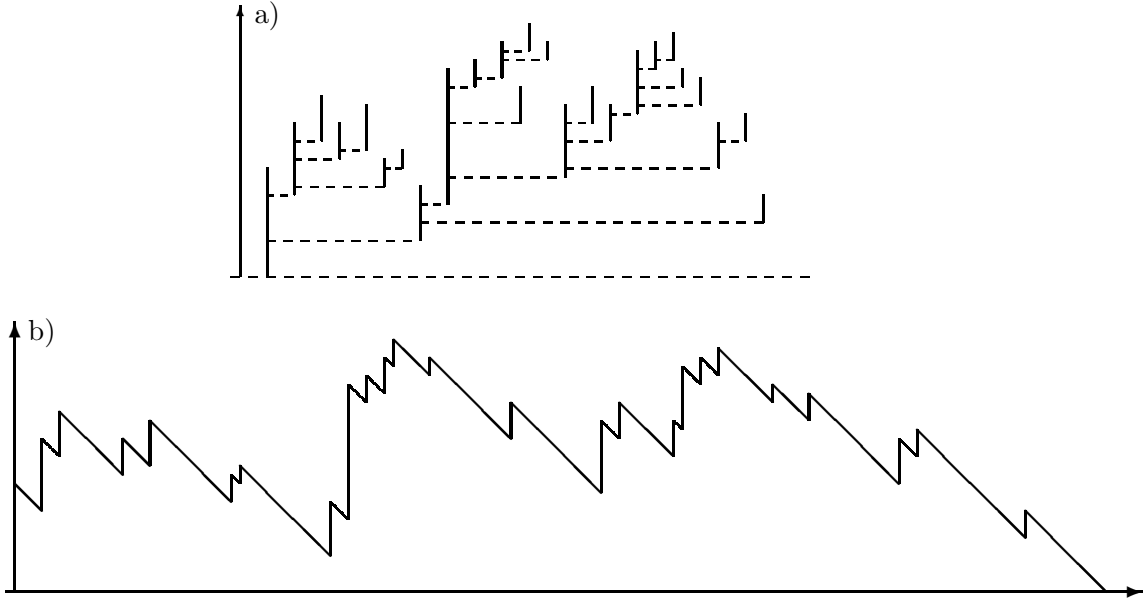
\begin{figure}[ht]
\unitlength 1.2mm 
\linethickness{0.4pt}
\ifx\plotpoint\undefined\newsavebox{\plotpoint}\fi 
\begin{picture}(129,70)(0,0)
\thicklines
\put(32,40){\line(0,1){12}}
\thinlines
\put(35,49){\line(1,0){.07}}
\put(31.93,48.93){\line(1,0){.75}}
\put(33.43,48.93){\line(1,0){.75}}
\put(0,0){}
\thicklines
\put(35,49){\line(0,1){8}}
\thinlines
\put(38,55){\line(1,0){.07}}
\put(34.93,54.93){\line(1,0){.75}}
\put(36.43,54.93){\line(1,0){.75}}
\put(0,0){}
\thicklines
\put(38,55){\line(0,1){5}}
\thinlines
\put(40,53){\line(1,0){.07}}
\put(34.93,52.93){\line(1,0){.833}}
\put(36.6,52.93){\line(1,0){.833}}
\put(38.26,52.93){\line(1,0){.833}}
\put(0,0){}
\thicklines
\put(40,53){\line(0,1){4}}
\put(43,54){\line(0,1){5}}
\put(49,44){\line(0,1){6}}
\thinlines
\put(52,48){\line(1,0){.07}}
\put(48.93,47.93){\line(1,0){.75}}
\put(50.43,47.93){\line(1,0){.75}}
\put(0,0){}
\thicklines
\put(52,48){\line(0,1){15}}
\thinlines
\put(55,61){\line(1,0){.07}}
\put(51.93,60.93){\line(1,0){.75}}
\put(53.43,60.93){\line(1,0){.75}}
\put(0,0){}
\thicklines
\put(55,61){\line(0,1){3}}
\thinlines
\put(43,54){\line(1,0){.07}}
\put(39.93,53.93){\line(1,0){.75}}
\put(41.43,53.93){\line(1,0){.75}}
\put(0,0){}
\put(58,62){\line(1,0){.07}}
\put(54.93,61.93){\line(1,0){.75}}
\put(56.43,61.93){\line(1,0){.75}}
\put(0,0){}
\thicklines
\put(58,62){\line(0,1){4}}
\thinlines
\put(61,65){\line(1,0){.07}}
\put(57.93,64.93){\line(1,0){.75}}
\put(59.43,64.93){\line(1,0){.75}}
\put(0,0){}
\thicklines
\put(61,65){\line(0,1){3}}
\thinlines
\put(63,64){\line(1,0){.07}}
\put(57.93,63.93){\line(1,0){.833}}
\put(59.6,63.93){\line(1,0){.833}}
\put(61.26,63.93){\line(1,0){.833}}
\put(0,0){}
\thicklines
\put(63,64){\line(0,1){2}}
\thinlines
\put(60,57){\line(1,0){.07}}
\put(51.93,56.93){\line(1,0){.889}}
\put(53.71,56.93){\line(1,0){.889}}
\put(55.49,56.93){\line(1,0){.889}}
\put(57.26,56.93){\line(1,0){.889}}
\put(59.04,56.93){\line(1,0){.889}}
\put(0,0){}
\thicklines
\put(60,57){\line(0,1){4}}
\thinlines
\put(65,51){\line(1,0){.07}}
\put(51.93,50.93){\line(1,0){.929}}
\put(53.79,50.93){\line(1,0){.929}}
\put(55.64,50.93){\line(1,0){.929}}
\put(57.5,50.93){\line(1,0){.929}}
\put(59.36,50.93){\line(1,0){.929}}
\put(61.22,50.93){\line(1,0){.929}}
\put(63.07,50.93){\line(1,0){.929}}
\put(0,0){}
\thicklines
\put(65,51){\line(0,1){8}}
\thinlines
\put(68,57){\line(1,0){.07}}
\put(64.93,56.93){\line(1,0){.75}}
\put(66.43,56.93){\line(1,0){.75}}
\put(0,0){}
\thicklines
\put(68,57){\line(0,1){4}}
\thinlines
\put(70,55){\line(1,0){.07}}
\put(64.93,54.93){\line(1,0){.833}}
\put(66.6,54.93){\line(1,0){.833}}
\put(68.26,54.93){\line(1,0){.833}}
\put(0,0){}
\thicklines
\put(70,55){\line(0,1){4}}
\thinlines
\put(73,58){\line(1,0){.07}}
\put(69.93,57.93){\line(1,0){.75}}
\put(71.43,57.93){\line(1,0){.75}}
\put(0,0){}
\thicklines
\put(73,58){\line(0,1){7}}
\thinlines
\put(78,61){\line(1,0){.07}}
\put(72.93,60.93){\line(1,0){.833}}
\put(74.6,60.93){\line(1,0){.833}}
\put(76.26,60.93){\line(1,0){.833}}
\put(0,0){}
\put(80,59){\line(1,0){.07}}
\put(72.93,58.93){\line(1,0){.875}}
\put(74.68,58.93){\line(1,0){.875}}
\put(76.43,58.93){\line(1,0){.875}}
\put(78.18,58.93){\line(1,0){.875}}
\put(0,0){}
\thicklines
\put(80,59){\line(0,1){3}}
\thinlines
\put(82,52){\line(1,0){.07}}
\put(64.93,51.93){\line(1,0){.944}}
\put(66.82,51.93){\line(1,0){.944}}
\put(68.71,51.93){\line(1,0){.944}}
\put(70.6,51.93){\line(1,0){.944}}
\put(72.49,51.93){\line(1,0){.944}}
\put(74.37,51.93){\line(1,0){.944}}
\put(76.26,51.93){\line(1,0){.944}}
\put(78.15,51.93){\line(1,0){.944}}
\put(80.04,51.93){\line(1,0){.944}}
\put(0,0){}
\thicklines
\put(82,52){\line(0,1){5}}
\thinlines
\put(85,55){\line(1,0){.07}}
\put(81.93,54.93){\line(1,0){.75}}
\put(83.43,54.93){\line(1,0){.75}}
\put(0,0){}
\thicklines
\put(85,55){\line(0,1){3}}
\thinlines
\put(87,46){\line(1,0){.07}}
\put(48.93,45.93){\line(1,0){.974}}
\put(50.88,45.93){\line(1,0){.974}}
\put(52.83,45.93){\line(1,0){.974}}
\put(54.78,45.93){\line(1,0){.974}}
\put(56.72,45.93){\line(1,0){.974}}
\put(58.67,45.93){\line(1,0){.974}}
\put(60.62,45.93){\line(1,0){.974}}
\put(62.57,45.93){\line(1,0){.974}}
\put(64.52,45.93){\line(1,0){.974}}
\put(66.47,45.93){\line(1,0){.974}}
\put(68.42,45.93){\line(1,0){.974}}
\put(70.37,45.93){\line(1,0){.974}}
\put(72.31,45.93){\line(1,0){.974}}
\put(74.26,45.93){\line(1,0){.974}}
\put(76.21,45.93){\line(1,0){.974}}
\put(78.16,45.93){\line(1,0){.974}}
\put(80.11,45.93){\line(1,0){.974}}
\put(82.06,45.93){\line(1,0){.974}}
\put(84.01,45.93){\line(1,0){.974}}
\put(85.96,45.93){\line(1,0){.974}}
\put(0,0){}
\thicklines
\put(87,46){\line(0,1){3}}
\thinlines
\put(49,44){\line(1,0){.07}}
\put(31.93,43.93){\line(1,0){.944}}
\put(33.82,43.93){\line(1,0){.944}}
\put(35.71,43.93){\line(1,0){.944}}
\put(37.6,43.93){\line(1,0){.944}}
\put(39.49,43.93){\line(1,0){.944}}
\put(41.37,43.93){\line(1,0){.944}}
\put(43.26,43.93){\line(1,0){.944}}
\put(45.15,43.93){\line(1,0){.944}}
\put(47.04,43.93){\line(1,0){.944}}
\put(0,0){}
\put(45,50){\line(1,0){.07}}
\put(34.93,49.93){\line(1,0){.909}}
\put(36.75,49.93){\line(1,0){.909}}
\put(38.57,49.93){\line(1,0){.909}}
\put(40.38,49.93){\line(1,0){.909}}
\put(42.2,49.93){\line(1,0){.909}}
\put(44.02,49.93){\line(1,0){.909}}
\put(0,0){}
\thicklines
\put(45,50){\line(0,1){3}}
\thinlines
\put(47,52){\line(1,0){.07}}
\put(44.93,51.93){\line(1,0){.667}}
\put(46.26,51.93){\line(1,0){.667}}
\put(0,0){}
\thicklines
\put(47,52){\line(0,1){2}}
\thinlines
\put(29,40){\vector(0,1){30}}
\put(27.941,39.941){\line(1,0){.9866}}
\put(29.915,39.941){\line(1,0){.9866}}
\put(31.888,39.941){\line(1,0){.9866}}
\put(33.861,39.941){\line(1,0){.9866}}
\put(35.834,39.941){\line(1,0){.9866}}
\put(37.808,39.941){\line(1,0){.9866}}
\put(39.781,39.941){\line(1,0){.9866}}
\put(41.754,39.941){\line(1,0){.9866}}
\put(43.727,39.941){\line(1,0){.9866}}
\put(45.701,39.941){\line(1,0){.9866}}
\put(47.674,39.941){\line(1,0){.9866}}
\put(49.647,39.941){\line(1,0){.9866}}
\put(51.62,39.941){\line(1,0){.9866}}
\put(53.593,39.941){\line(1,0){.9866}}
\put(55.567,39.941){\line(1,0){.9866}}
\put(57.54,39.941){\line(1,0){.9866}}
\put(59.513,39.941){\line(1,0){.9866}}
\put(61.486,39.941){\line(1,0){.9866}}
\put(63.46,39.941){\line(1,0){.9866}}
\put(65.433,39.941){\line(1,0){.9866}}
\put(67.406,39.941){\line(1,0){.9866}}
\put(69.379,39.941){\line(1,0){.9866}}
\put(71.353,39.941){\line(1,0){.9866}}
\put(73.326,39.941){\line(1,0){.9866}}
\put(75.299,39.941){\line(1,0){.9866}}
\put(77.272,39.941){\line(1,0){.9866}}
\put(79.245,39.941){\line(1,0){.9866}}
\put(81.219,39.941){\line(1,0){.9866}}
\put(83.192,39.941){\line(1,0){.9866}}
\put(85.165,39.941){\line(1,0){.9866}}
\put(87.138,39.941){\line(1,0){.9866}}
\put(89.112,39.941){\line(1,0){.9866}}
\put(91.085,39.941){\line(1,0){.9866}}
\thicklines
\put(78,61){\line(0,1){2}}
\thinlines
\put(75,63){\line(1,0){.07}}
\put(72.93,62.93){\line(1,0){.667}}
\put(74.26,62.93){\line(1,0){.667}}
\put(0,0){}
\thicklines
\put(75,63){\line(0,1){3}}
\thinlines
\put(77,64){\line(1,0){.07}}
\put(74.93,63.93){\line(1,0){.667}}
\put(76.26,63.93){\line(1,0){.667}}
\put(0,0){}
\thicklines
\put(77,64){\line(0,1){3}}
\put(4,5){\vector(0,1){30}}
\put(4,17){\line(1,-1){3}}
\multiput(7,22)(.02777778,-.02777778){72}{\line(0,-1){.02777778}}
\put(9,25){\line(1,-1){7}}
\put(16,22){\line(1,-1){3}}
\put(19,24){\line(1,-1){9}}
\multiput(28,18)(.02777778,-.02777778){36}{\line(0,-1){.02777778}}
\put(29,19){\line(1,-1){10}}
\multiput(39,15)(.02777778,-.02777778){72}{\line(0,-1){.02777778}}
\multiput(41,28)(.02777778,-.02777778){72}{\line(0,-1){.02777778}}
\multiput(43,29)(.02777778,-.02777778){72}{\line(0,-1){.02777778}}
\put(45,27){\line(0,1){4}}
\multiput(45,31)(.02777778,-.02777778){36}{\line(0,-1){.02777778}}
\put(46,30){\line(0,1){3}}
\put(46,33){\line(1,-1){4}}
\put(50,29){\line(0,1){2}}
\put(50,31){\line(1,-1){9}}
\put(59,22){\line(0,1){4}}
\put(59,26){\line(1,-1){10}}
\put(69,16){\line(0,1){8}}
\multiput(69,24)(.02777778,-.02777778){72}{\line(0,-1){.02777778}}
\put(71,22){\line(0,1){4}}
\put(71,26){\line(1,-1){6}}
\put(77,20){\line(0,1){4}}
\multiput(77,24)(.02777778,-.02777778){36}{\line(0,-1){.02777778}}
\put(78,23){\line(0,1){7}}
\multiput(78,30)(.02777778,-.02777778){72}{\line(0,-1){.02777778}}
\put(80,28){\line(0,1){3}}
\multiput(80,31)(.02777778,-.02777778){72}{\line(0,-1){.02777778}}
\put(82,29){\line(0,1){3}}
\put(82,32){\line(1,-1){6}}
\put(88,26){\line(0,1){2}}
\put(88,28){\line(1,-1){4}}
\put(92,24){\line(0,1){3}}
\put(92,27){\line(1,-1){10}}
\put(102,17){\line(0,1){5}}
\multiput(102,22)(.02777778,-.02777778){72}{\line(0,-1){.02777778}}
\put(104,20){\line(0,1){3}}
\put(104,23){\line(1,-1){12}}
\put(116,11){\line(0,1){3}}
\put(116,14){\line(1,-1){9}}
\put(3,5){\vector(1,0){126}}
\put(7,14){\line(0,1){8}}
\put(9,20){\line(0,1){5}}
\put(16,18){\line(0,1){4}}
\put(19,19){\line(0,1){5}}
\put(28,15){\line(0,1){3}}
\put(29,17){\line(0,1){2}}
\put(39,9){\line(0,1){6}}
\put(41,13){\line(0,1){15}}
\put(43,26){\line(0,1){3}}
\put(32,68.875){\makebox(0,0)[cc]{a)}}
\put(7,33.75){\makebox(0,0)[cc]{b)}}
\end{picture}

\caption{a) A realization of a \emph{splitting tree} with finite extinction time. Horizontal axis has no interpretation, but horizontal arrows indicate filiation; vertical axis indicates real time; b) The associated jumping chronological \emph{contour process} with jumps in solid line. }
\label{fig : jccp}
\end{figure}

In \cite{L}, we have considered the so-called jumping chronological contour process (JCCP) of the splitting tree truncated up to height (time) $t$, which starts at $\min (s,t)$, where $s$ is the death time of the progenitor, visits all existence times (smaller than $t$) of all individuals exactly once and terminates at 0.
We have shown \cite[Theorem 4.3]{L} that the JCCP is a Markov process, more specifically, it is a compound Poisson process $X$ with jump measure $\pi$, compensated at rate $-1$, reflected below $t$, and killed upon hitting 0. We denote the law of $X$ by $P$, to make the difference with the law $\PP$ of the CMJ process. As seen previously, we record the lifetime duration, say $s$, of the progenitor, by writing $P_s$ for its conditional law on $X_0=s$.

Let us be a little more specific about the JCCP. Recall that this process visits all existence times of all individuals of the truncated tree. For any individual of the tree, we denote by $\alpha$ its birth time and by $\omega$ its death time. When the visit of an individual $v$ with lifespan $(\alpha(v), \omega(v)]$ begins, the value of the JCCP is $\omega(v)$. The JCCP then visits all the existence times of $v$'s lifespan at constant speed $-1$. If $v$ has no child, then this visit lasts exactly the lifespan of $v$; if $v$ has at least one child, then the visit is interrupted each time a birth time of one of $v$'s daughters, say $w$, is encountered (youngest child first since the visit started at the death level). At this point, the JCCP jumps from $\alpha(w)$ to $\omega(w)\wedge t$ and starts the visit of the existence times of $w$. Since the tree has finite length, the visit of $v$ has to terminate: it does so at the chronological level $\alpha(v)$ and continues the exploration of the existence times of $v$'s mother, at the height (time) where it had been interrupted.
This procedure then goes on recursively as soon as $0$ is encountered (birth time of the progenitor). See Figure \ref{fig : jccp} for an example.

\begin{figure}[ht]

\unitlength 1.8mm 
\linethickness{0.4pt}
\ifx\plotpoint\undefined\newsavebox{\plotpoint}\fi 
\begin{picture}(69,37)(-4,3)
\put(6,5){\vector(0,1){35}}
\put(4.961,5.961){\line(1,0){.9836}}
\put(6.928,5.961){\line(1,0){.9836}}
\put(8.895,5.961){\line(1,0){.9836}}
\put(10.863,5.961){\line(1,0){.9836}}
\put(12.83,5.961){\line(1,0){.9836}}
\put(14.797,5.961){\line(1,0){.9836}}
\put(16.764,5.961){\line(1,0){.9836}}
\put(18.731,5.961){\line(1,0){.9836}}
\put(20.699,5.961){\line(1,0){.9836}}
\put(22.666,5.961){\line(1,0){.9836}}
\put(24.633,5.961){\line(1,0){.9836}}
\put(26.6,5.961){\line(1,0){.9836}}
\put(28.568,5.961){\line(1,0){.9836}}
\put(30.535,5.961){\line(1,0){.9836}}
\put(32.502,5.961){\line(1,0){.9836}}
\put(34.469,5.961){\line(1,0){.9836}}
\put(36.436,5.961){\line(1,0){.9836}}
\put(38.404,5.961){\line(1,0){.9836}}
\put(40.371,5.961){\line(1,0){.9836}}
\put(42.338,5.961){\line(1,0){.9836}}
\put(44.305,5.961){\line(1,0){.9836}}
\put(46.272,5.961){\line(1,0){.9836}}
\put(48.24,5.961){\line(1,0){.9836}}
\put(50.207,5.961){\line(1,0){.9836}}
\put(52.174,5.961){\line(1,0){.9836}}
\put(54.141,5.961){\line(1,0){.9836}}
\put(56.108,5.961){\line(1,0){.9836}}
\put(58.076,5.961){\line(1,0){.9836}}
\put(60.043,5.961){\line(1,0){.9836}}
\put(62.01,5.961){\line(1,0){.9836}}
\put(63.977,5.961){\line(1,0){.9836}}
\put(10,20){\line(0,-1){14}}
\put(16,28){\line(0,-1){10}}
\put(22,23){\line(0,1){8}}
\put(27,29){\line(0,1){7}}
\put(36,26){\line(0,1){12}}
\put(44,11){\line(0,1){9}}
\put(50,16){\line(0,1){19}}
\put(15.961,17.961){\line(-1,0){.8571}}
\put(14.247,17.961){\line(-1,0){.8571}}
\put(12.532,17.961){\line(-1,0){.8571}}
\put(10.818,17.961){\line(-1,0){.8571}}
\put(21.961,22.961){\line(-1,0){.8571}}
\put(20.247,22.961){\line(-1,0){.8571}}
\put(18.532,22.961){\line(-1,0){.8571}}
\put(16.818,22.961){\line(-1,0){.8571}}
\put(26.961,28.961){\line(-1,0){.8333}}
\put(25.294,28.961){\line(-1,0){.8333}}
\put(23.628,28.961){\line(-1,0){.8333}}
\put(35.961,25.961){\line(-1,0){.9333}}
\put(34.094,25.961){\line(-1,0){.9333}}
\put(32.228,25.961){\line(-1,0){.9333}}
\put(30.361,25.961){\line(-1,0){.9333}}
\put(28.494,25.961){\line(-1,0){.9333}}
\put(26.628,25.961){\line(-1,0){.9333}}
\put(24.761,25.961){\line(-1,0){.9333}}
\put(22.894,25.961){\line(-1,0){.9333}}
\put(49.961,15.961){\line(-1,0){.8571}}
\put(48.247,15.961){\line(-1,0){.8571}}
\put(46.532,15.961){\line(-1,0){.8571}}
\put(44.818,15.961){\line(-1,0){.8571}}
\put(43.961,10.961){\line(-1,0){.9962}}
\put(41.969,10.961){\line(-1,0){.9962}}
\put(39.976,10.961){\line(-1,0){.9962}}
\put(37.984,10.961){\line(-1,0){.9962}}
\put(35.992,10.961){\line(-1,0){.9962}}
\put(33.999,10.961){\line(-1,0){.9962}}
\put(32.007,10.961){\line(-1,0){.9962}}
\put(30.014,10.961){\line(-1,0){.9962}}
\put(28.022,10.961){\line(-1,0){.9962}}
\put(26.03,10.961){\line(-1,0){.9962}}
\put(24.037,10.961){\line(-1,0){.9962}}
\put(22.045,10.961){\line(-1,0){.9962}}
\put(20.053,10.961){\line(-1,0){.9962}}
\put(18.06,10.961){\line(-1,0){.9962}}
\put(16.068,10.961){\line(-1,0){.9962}}
\put(14.076,10.961){\line(-1,0){.9962}}
\put(12.083,10.961){\line(-1,0){.9962}}
\multiput(4.961,32.961)(.983333,0){61}{{\rule{.4pt}{.4pt}}}
\put(58,23){\line(0,1){5}}
\put(64,25){\line(0,1){11}}
\put(57,12){\line(0,1){4}}
\put(56.961,11.961){\line(-1,0){.9286}}
\put(55.104,11.961){\line(-1,0){.9286}}
\put(53.247,11.961){\line(-1,0){.9286}}
\put(51.39,11.961){\line(-1,0){.9286}}
\put(49.532,11.961){\line(-1,0){.9286}}
\put(47.675,11.961){\line(-1,0){.9286}}
\put(45.818,11.961){\line(-1,0){.9286}}
\put(57.961,22.961){\line(-1,0){.8889}}
\put(56.183,22.961){\line(-1,0){.8889}}
\put(54.405,22.961){\line(-1,0){.8889}}
\put(52.628,22.961){\line(-1,0){.8889}}
\put(50.85,22.961){\line(-1,0){.8889}}
\put(63.961,24.961){\line(-1,0){.8571}}
\put(62.247,24.961){\line(-1,0){.8571}}
\put(60.532,24.961){\line(-1,0){.8571}}
\put(58.818,24.961){\line(-1,0){.8571}}
\thicklines
\put(54,33){\vector(0,-1){10}}
\put(31.5,33){\vector(0,-1){7}}
\put(40,33){\vector(0,-1){22}}
\put(4,33){\makebox(0,0)[cc]{$t$}}
\put(25,34){\makebox(0,0)[cc]{$x_1$}}
\put(34,34){\makebox(0,0)[cc]{$x_2$}}
\put(48,34){\makebox(0,0)[cc]{$x_3$}}
\put(62,34){\makebox(0,0)[cc]{$x_4$}}
\put(33,29){\makebox(0,0)[cc]{$H_1$}}
\put(41.5,24){\makebox(0,0)[cc]{$H_2$}}
\put(55.5,29){\makebox(0,0)[cc]{$H_3$}}
\thinlines
\put(26.588,32.957){\line(1,0){.946}}
\put(35.522,32.957){\line(1,0){1.051}}
\put(49.502,32.957){\line(1,0){1.051}}
\put(63.482,32.957){\line(1,0){1.051}}
\end{picture}

\caption{Illustration of a splitting tree showing the durations $H_1, H_2, H_3$ elapsed since \emph{coalescence} for each of the three consecutive pairs $(x_1, x_2), (x_2, x_3)$ and $(x_3, x_4)$ of the $N_t=4$ individuals alive at time $t$.
}
\label{fig : treecoal}
\end{figure}
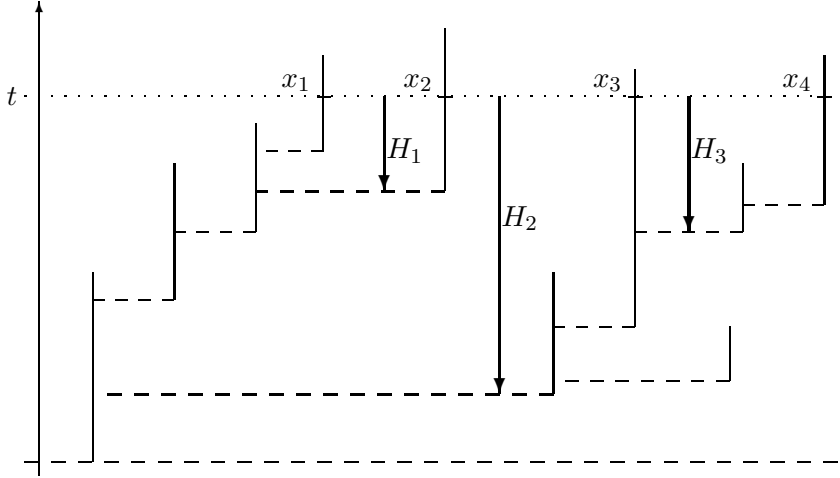

Since the JCCP is Markovian (as seen earlier, it is a reflected, killed Lévy process), its excursions between consecutive visits of points at height $t$ are i.i.d. excursions of $X$. Observe in particular that the number of visits of $t$ by $X$ is exactly the number $N_t$ of individuals alive at time $t$, where $N$ is the aforementioned homogeneous, binary Crump--Mode--Jagers process. See Figure \ref{fig : treecoal}.

This property has two consequences, the first of which will be exploited in the immigration model, and the second one in the mutation model.

The first consequence is the computation of the one-dimensional marginals of $N$. Let $T_A$ denote the first hitting time of the set $A$ by $X$. Conditional on the initial progenitor to have lived $s$ units of time, we have
\begin{equation}
\label{eqn : N_t=0}
\PP_s(N_t=0)=P_s(T_0< T_{(t,+\infty)}),
\end{equation}
and, applying recursively the strong Markov property,
\begin{equation}
\label{eqn : N_t=k condition}
\PP_s(N_t=k \mid N_t\not=0) = P_t(T_{(t,+\infty)}<T_0)^{k-1}P_t(T_0< T_{(t,+\infty)}).
\end{equation}
Note that the subscript $s$ in the last display is useless.

The second consequence is that because $X$ is (strongly) Markovian, the depths of the excursions of $X$ away from $t$ are i.i.d., distributed as some random variable $H:=t-\inf_{0\le s\le T} X_s$, where $X$ is started at $t$ and $T$ denotes the first hitting time $T_0\wedge T_{(t,+\infty)}$ of $\{0\}\cup(t,+\infty)$ by $X$. We record this by letting $H_i$ denote the depth of the excursion between the $i$-th visit of $t$ and its $(i+1)$-th visit, and stating that the variables $H_1, H_2,\ldots$ form a sequence of i.i.d. random variables distributed as $H$ and killed at its first value greater than $t$.

But in the splitting tree,  $H_i$ is also the \emph{coalescence time} (or \emph{divergence time}) between individual $i$ and individual $i+1$, that is, the time elapsed since the lineages of individual $i$ and $i+1$ have diverged. Further, it can actually be shown \cite{L} that the coalescence time $C_{i,i+k}$ between individual $i$ and individual $i+k$ is given by
\begin{equation}
\label{eqn : def coal}
C_{i,i+k}=\max\{H_{i+1},\ldots,H_{i+k}\}, 
\end{equation}
so that the genealogical structure of the alive population of a splitting tree is entirely given by the knowledge of a sequence of independent random variables
$H_1, H_2,\ldots$ that we will call \emph{branch lengths}, all distributed as $H$. We call the whole sequence the \emph{coalescent point process}. 

Here, exact formulae can be deduced for \eqref{eqn : N_t=0} and \eqref{eqn : N_t=k condition} from the fact that the JCCP is a Lévy process with no negative jumps. In particular, it can be convenient to handle its Laplace exponent $\psi$ instead of its jump measure $\pi$, that is,
\begin{equation}
\label{eqn : psi}
\psi(a):= a -\intgen \pi(dx) (1-e^{-a x}) \qquad a\ge 0.
\end{equation}
We know \cite{L} that the process is subcritical, critical or supercritical, according to whether $m:=\int_{(0,\infty]} r\pi(dr)<1$, $=1$ or $>1$. In the latter case, the rate $\eta$ at which $(N_t;t\ge 0)$ grows exponentially on the  event of non-extinction, called the \emph{Malthusian parameter},  is the only nonzero root of the convex function $\psi$.
Furthermore, the probability of exit of an interval (from the bottom or from the top) by $X$  has a simple expression (see e.g. \cite{B}), in the form
\begin{equation}
\label{eqn : two-sided}
P_s(T_0< T_{(t,+\infty)}) = \frac{W(t-s)}{W(t)},
\end{equation}
where the so-called \emph{scale function} $W$ is the nonnegative, nondecreasing, differentiable function such that $W(0)=1$, characterized by its Laplace transform 
\begin{equation}
\label{eqn : LT scale}
\intgen dx\, e^{-a x} \, W(x) = \frac{1}{\psi(a)} \qquad a>\eta.
\end{equation}
As a consequence, the typical branch length $H$ between two consecutive individuals alive at time $t$ has the following distribution (conditional on there being at least two extant individuals at time $t$)
\begin{equation}
\label{eqn : conditional height}
\PP(H< s) =  P_t(T_{(t,+\infty)}<T_s\mid  T_{(t,+\infty)}<T_0) =\frac{1-\frac{1}{W(s)}}{1-\frac{1}{W(t)}}\qquad 0\le s\le t.
\end{equation}
Let us stress that in some examples,  \eqref{eqn : LT scale} can be inverted.
When $\pi$ has an exponential density, $(N_t;t\ge 0)$ is a linear birth--death process with (birth rate $\lbd$ and) death rate, say $\rho$. If $\lbd\not=\rho$, then (see \cite{L} for example)
$$
W(x) = \frac{\rho-\lbd e^{(\lbd-\rho)x}}{\rho-\lbd}\qquad x\ge 0,
$$
whereas if $\lbd=\rho$,
$$
W(x)=1+\lbd x\qquad x\ge 0.
$$
When $\pi$ is a point mass at $\infty$, $(N_t;t\ge0)$ is a pure-birth process, called Yule process, with birth rate $\lbd$. Then (let $\rho\to 0$)
$$
W(x)=e^{\lbd x}\qquad x\ge 0.
$$
In the case when $\lambda\not=\rho\not=0$, it had already been noticed by B. Rannala \cite{Rannala97} that the coalescence times of a population whose genealogy is given by a (linear) birth--death process started (singly) $t$ units of time ago and whose size is conditioned to be $n$, are identical to those of the order statistics of $n$ i.i.d. random variables with density
$$
f(s)=\frac{(1-p_0(s))(\rho-\lambda p_0(s))}{p_0(t)}\qquad 0< s< t,
$$
where $\rho$ is the death rate and
$$
p_0(t) := \frac{\rho\left(e^{rt}-1\right)}{\lambda e^{rt}-\rho} ,
$$
where $r:=\lambda -\rho$.
Now \eqref{eqn : conditional height} applied to the expression of the scale function given previously for the birth--death case ($\lambda\not=\rho$) agrees with the findings of B. Rannala under the form
$$
f(s)\ ds=\PP(H\in ds) = \frac{r^2 \ e^{rs}}{\left(\lambda e^{rs} -\rho\right)^2}\cdot\frac{
\lambda e^{rt}-\rho}{e^{rt}-1}\ ds \qquad 0<s<t.
$$
It is remarkable that in this case, exchanging $\lambda$ and $\rho$ leaves the distribution of $H$ unchanged. No extension of this fact is known in the general case.

We end this section by the following lemma.
\begin{lem}
\label{lem : one-dim N}
The one-dimensional marginal of $N_t$ when the lifespan of the progenitor is random with law $\pi(\cdot)/\lambda$, is given by
$$
\PP(N_t\not=0)= \frac{W'(t)}{\lbd W(t)}\qquad t\ge 0
$$
and 
$$
\PP(N_t=k) =\left(1-\frac{1}{W(t)}\right)^{k-1} \frac{W'(t)}{\lbd W(t)^2}\qquad t\ge 0.
$$
\end{lem}
\paragraph{Proof.}
From \eqref{eqn : N_t=0} and \eqref{eqn : two-sided},  we get
$$
\PP_s(N_t=0)=\frac{W(t-s)}{W(t)}
$$
and from \eqref{eqn : N_t=k condition} and \eqref{eqn : two-sided}, we get
$$
\PP_s(N_t=k \mid N_t\not=0) = \left(1-\frac{1}{W(t)}\right)^{k-1}\frac{1}{W(t)}.
$$
Let us compute the unconditional law of $N_t$ by integrating over $s$. First,
$$
\PP(N_t=0)=\int_0^t \lbd^{-1}\pi(ds)\frac{W(t-s)}{W(t)}= \frac{F(t)}{\lbd W(t)},
$$
where
$$
F(t):= \int_0^t \pi(ds)W(t-s) \qquad t\ge 0.
$$
Now by Fubini--Tonelli,
$$
\intgen dt\, F(t) e^{-at} = \intgen \pi(ds)\int_s^\infty dt e^{-at} W(t-s)=\frac{1}{\psi(a)}\intgen \pi(ds) e^{-as},
$$
referring to \eqref{eqn : LT scale}, where we recall from \eqref{eqn : psi} that
$$
\psi(a)= a -\intgen \pi(dx) (1-e^{-a x})=a-\lbd + \intgen \pi(dx) e^{-a x} \qquad a\ge 0.
$$
This yields
$$
\intgen dt\, F(t) e^{-at} = 1+\frac{\lbd -a}{\psi(a)}.
$$
This Laplace transform can be inverted as follows
$$
F(t)=\lbd W(t)-W'(t)\qquad t\ge 0.
$$
Thus, we get the announced expression for $N_t$. \hfill $\Box$

\section{The immigration model}

Assume that we start at time $0$ on the island with no individual at all. Let $I_t$ denote the total number of extant individuals at time $t$. Let $I_t(k)$ denote the number of species (each corresponding to a single progenitor immigrant) with $k$ representative individuals at time $t$. In particular,
$$
I_t=\sum_{k\ge 1}k I_t(k).
$$
We allow $k$ to equal 0, $I_t(0)$ corresponding to the number of effective immigrants having 0 descendance at time $t$. Recall from the Preliminaries the scale function $W$.
\begin{thm}
\label{thm : spectrum mig} 
The random variables $(I_t(0), I_t(1), \ldots)$ are independent Poisson random variables. For any $k\not=0$, the r.v. $I_t(k)$ is a Poisson r.v. with parameter
$$
\frac{\gamma}{ k } \left(1-\frac{1}{ W(t)}\right)^k,
$$
where $\gamma:=\mu/\lbd$ is the immigration-to-birth ratio. The Poisson r.v. $I_t(0)$ has parameter
$$
\mu t -\gamma \ln W(t).
$$
\end{thm}
Thanks to a standard result on independent Poisson random variables $X_k$ with respective means $c\alpha^k/k$, conditioning on $\sum kX_k$ removes the dependence in $\alpha$ (see e.g. \cite[p.220]{W}). It is then remarkable that conditioning the frequency spectrum on the total number of individuals \emph{removes the dependence in $t$}. In the case of exponential lifetimes, this property has been re-discovered various times, see for example \cite{Rannala96}. Here, the conditioning does \emph{not only} remove the dependence in $t$, but also in $\lambda$, $W$ or $\pi$, that is, \emph{in the whole dynamical scheme distribution}. 
\begin{cor}
\label{cor : Poisson cond}
 Let $Y_1, Y_2, \ldots$ be independent random variables, where $Y_k$ follows the Poisson distribution with parameter $\gamma/ k$. 
Conditional on the total number $I_t$ of species at time $t$ equalling $n$, the random vector $(I_t(1), \ldots, I_t(n))$, then also  denoted $(I_n(1), \ldots, I_n(n))$, has the same law as $(Y_1,\ldots, Y_n)$ conditioned by $\sum_{k=1}^n kY_k = n$.
\end{cor}
\begin{rem}
This conditional spectrum is exactly the same one as that obtained in the Kingman coalescent with mutations at rate $\gamma$ in the infinite-alleles model (i.e., the spectrum given by Ewens' sampling formula). In the case of exponential lifetimes, this coincidence between the binary branching process with immigration and the Moran process with mutations can be explained thanks to Hoppe's urn model (see \cite{D}). This observation has been recast in the neutral theory of biodiversity literature as a possible relaxation of the `zero-sum assumption' \cite{EAMcK,HE}. 
\end{rem}
\begin{rem}
Theorem \ref{thm : spectrum mig} is concerned with species with fixed abundances $k=1,2,\ldots$, i.e., the `small' families. It is also possible to get results for the abundances $P_1, P_2,\ldots$ of the immmigrant surviving families ranked by decreasing order of ages, i.e., the `large' families, either as the population size $n\to\infty$ or as time $t\to\infty$ in the supercritical case (mean number of offspring $m>1$). M. Richard \cite{Richard} obtains that the vector $(P_1,P_2,\cdots)$ rescaled by population size converges a.s. to the GEM distribution with parameter $\gamma$.
\end{rem}
Let us now prove the theorem. Let $M_t$ be the number of immigrants having reached the island up until time $t$, and $T_1<\cdots <T_{M_t} <t$ the times of arrival of these immigrants. For any integer $n$, let $\sigma_n$ denote an independent, random (uniform) permutation on $\{1,\ldots, n\}$.
Then $M_t$ is a Poisson r.v. with parameter $\mu t$, and conditional on  $M_t=n$, the random variables $(T_{\sigma_n(1)},\ldots, T_{\sigma_n(n)})$ are i.i.d., uniformly distributed on $[0,t]$. Then we call $Z_t^{(i)}$ the number of descendants at time $t$ of the particle having immigrated at time $T_{\sigma_n(i)}$. The random variables $(Z_t^{(i)}, i=1,\ldots,n)$ are i.i.d. distributed as some r.v. $Z_t$ which is the value of the Crump--Mode--Jagers process $N_t$ started at a uniform time on $[0,t]$
$$
\PP(Z_t^{(i)}=k)=\frac{1}{t}\int_0^t du \PP(N_{u}=k),
$$
where it will always be understood that $N_0=1$. 
The following statement is the key result to the theorem. 
\begin{prop} The law of $Z_t$ is given by the following two equations.
$$
\PP(Z_t=k)=\frac{1}{\lbd k t} \left(1-\frac{1}{ W(t)}\right)^k
$$
for $k\not=0$, whereas
$$
\PP(Z_t=0)=1-\frac{1}{\lbd t} \ln W(t).
$$
\end{prop}
Before proving the proposition, we remind the reader of an elementary lemma on multinomial distributions with Poisson randomizing parameter. The theorem follows from this lemma and the proposition.
\begin{lem}
Let $p:=(p_0, p_1,\ldots)$ be some probability distribution on the integers, let $X_1, X_2,\ldots$ be i.i.d. r.v. with law $p$ and let $B$ be an independent Poisson r.v. with parameter $\beta$. Finally, set
$$
B_k:=\#\{i=1,\ldots, B : X_i=k\} \qquad k\ge 0.
$$
Then the random variables $B_0, B_1,\ldots$ are independent Poisson r.v., and $B_k$ has parameter $\beta p_k$.
\end{lem}
\paragraph{Proof of the proposition.}
Thanks to Lemma \ref{lem : one-dim N}, we have
$$
\PP(N_t\not=0)= \frac{W'(t)}{\lbd W(t)}\qquad t\ge 0,
$$
and 
$$
\PP(N_t=k) =\left(1-\frac{1}{W(t)}\right)^{k-1} \frac{W'(t)}{\lbd W(t)^2}\qquad t\ge 0.
$$
Let us now turn to $Z_t$, which has the law of $N_t$ with origination time uniform on $[0,t]$. First,
$$
\PP(Z_t\not=0)= \frac{1}{t}\int_0^t du \PP(N_{u}\not=0) = \frac{1}{t}\int_0^t du \frac{W'(u)}{\lbd W(u)} =\frac{1}{\lbd t} \ln W(t).
$$
Second,
$$
\PP(Z_t=k)= \frac{1}{t}\int_0^t du \PP(N_{u}=k) = \frac{1}{t}\int_0^t du \left(1-\frac{1}{W(u)}\right)^{k-1} \frac{W'(u)}{\lbd W(u)^2}=\frac{1}{\lbd k t} \left(1-\frac{1}{ W(t)}\right)^k,
$$
which ends the proof of the proposition.\hfill $\Box$

\section{The mutation model}

Recall from the section on splitting trees and coalescent point processes that the genealogy at a fixed time $t$  of the $N_t$ extant individuals of the splitting tree, originating from a  single progenitor individual born at time 0, is characterized by the branch lengths $H_i$, $i=1,\ldots N_t-1$, where $H_i$ is the divergence time between individual $i$ and individual $i+1$. In addition, these r.v. are i.i.d. with common distribution
$$
\PP(H< s)  =\frac{1-\frac{1}{W(s)}}{1-\frac{1}{W(t)}}\qquad 0\le s\le t,
$$
where the so-called scale function $W$ depends on the birth rate $\lambda$ and on the lifespan measure $\pi$, and is characterized by its Laplace transform.

In the critical or supercritical cases, where $W$ is unbounded, we can define the long-lived tree asymptotics, by letting  $t\tendinfty$. This leads to
$$
\PP(H<s) =1-\frac{1}{{W(s)}}\qquad s\ge 0,
$$
and the \emph{stationary} genealogy is then given by an infinite sequence of branches with i.i.d. lengths, with tail as in the last display. In the subcritical case, $W$ has a finite limit equal to $1/(1-m)$ (see \cite{L}). Then conditioning on the population being still extant at time $t$ and letting $t\to\infty$, the \emph{quasi-stationary} genealogy is given by a parameter $m$ geometric number of branches with i.i.d. lengths distributed as follows 
$$
\PP^\star(H< s) =  m^{-1}\left(1-\frac{1}{{W(s)}}\right)\qquad s\ge 0,
$$
where the star superscript serves to remind the conditioning.

In this section, individuals experience mutations at rate $\theta$ during their lifetime, and each mutation yields a brand new type. This assumption corresponds to what is usually called the \emph{infinitely-many alleles model}. We now introduce the function $W_\theta$, which is the scale function associated to the so-called \emph{clonal process}. More specifically, if one restricts the tree to points bearing the same type (e.g., the same type as the progenitor's type), then one retrieves a new splitting tree, whose birth rate remains equal to $\lambda$ and whose lifetime durations are distributed as a r.v. $V^\theta$ defined as the minimum of $V$ and of an independent exponential variable with parameter $\theta$ (i.e., the first mutation event). As in \cite{L2}, we can then define $H^\theta$ as the divergence time between consecutive individuals in the clonal splitting tree. In the (more general) coalescent point process, $H^\theta$ is defined as the divergence time between individual 0 and the first individual whose type satisfies the following property: it is one of the successive types that appeared across time in the history of the lineage of individual 0. We have proved \cite{L2} that the function $W_\theta$ (either defined as the scale function of the clonal splitting tree or equivalently,  in the coalescent point process, as the inverse of the tail of $H^\theta$) satisfies
\begin{equation}
W_\theta(x)=1+\int_0^x W'(s)e^{-\theta s}\, ds\qquad x\ge 0.
\end{equation}
Now consider the standing population at time $t$ conditioned on being nonempty, whose probability law we denote by $\PP^\star$.
For any real number $y\in(0,t)$, define $A_t (k;dy)$ as the number of species originating in a point mutation having occurred during the time interval $(y, y+dy)$ and represented by exactly $k$ alive individuals at time $t$. The following proposition gives the expectation under $\PP^\star$ of $A_t(k;dy)$ and is extracted from \cite{CL1}.
\begin{thm}
\label{thm : exp freq spectrum}
For any $k\ge 1$, the expected number of species of age in $dy$ and abundance $k$ is
$$
\EE^\star A_t (k;dy) =\theta\, dy\, W(t)\frac{e^{-\theta y}}{W_\theta(y)^2}\left(1-\frac{1}{W_\theta(y)}\right)^{k-1} .
$$
\end{thm}
In \cite{CL1}, we provide arguments giving an intuition of this result. To be more specific, the last expression can be seen as the product of the three following terms :
$$
\theta \,dy\ \frac{W(t)}{W(y)}
$$
which is the sum over $i=1,2\ldots$ of the probabilities that the $i$-th branch length has size $H_i \ge y$ and (is the one that) carries a mutation with age in  $(y, y+dy)$, multiplied by 
$$
\frac{W(y)\,e^{-\theta y}}{W_\theta(y)}
$$
which is the probability that the type carried by the lineage of the $i$-th individual at time $t-y$ has at least one alive representative, finally multiplied by
$$
\frac{1}{W_\theta(y)}\left(1-\frac{1}{W_\theta(y)}\right)^{k-1} 
$$
which is the probability that the type carried by the lineage of the $i$-th individual at time $t-y$ has exactly $k$ alive representatives, conditional on having at least 1.\\

Recall that $A_t$ denotes the number of species in the population at time $t$ and that $A_t(k)$ denotes the number of species represented by exactly $k$ extant individuals.
We can record the last theorem under its integral representation :
\begin{prop}  For any $k\ge 1$,
$$
\EE^\star A_t(k)	=	W(t) \int_0^t dy \,\theta\,e^{-\theta y} \frac{1}{W_\theta(y)^2}\left(1-\frac{1}{W_\theta(y)}\right)^{k-1}
$$
and
$$
\EE^\star A_t	=	W(t) \int_0^t dy \,\theta\,e^{-\theta y} \frac{1}{W_\theta(y)} .
$$
\end{prop}
Furthermore, we got the following asymptotic result, extracted from \cite{CL1} and \cite{L2}. Here, $A_n(k)$ denotes the number of species with $k$ individuals in the coalescent point process with population size $n$. Recall that coalescent point processes with different population sizes can be constructed on the same space by merely adding new independent branches. This allows us to state pathwise convergences for $A_n$ as $n\to \infty$.
\begin{thm}
\label{thm : spectrum mut}
For all $k\ge 1$, the following convergence holds a.s., as $n\to\infty$ for the coalescent point process, and as $t\to\infty$ for the splitting tree in the supercritical case and on the event of non-extinction :
$$
\lim_{n\tendinfty} n^{-1}A_n(k) 	=\lim_{t\tendinfty} N_t^{-1}A_t(k)=	\intgen dy \,\theta\,e^{-\theta y} \frac{1}{W_\theta(y)^2}\left(1-\frac{1}{W_\theta(y)}\right)^{k-1}
$$
and
$$
\lim_{n\tendinfty} n^{-1}A_n 	=	\lim_{n\tendinfty} N_t^{-1}A_t(k)=\intgen dy \,\theta\,e^{-\theta y} \frac{1}{W_\theta(y)}.
$$
\end{thm}
\begin{rem}
The a.s. result for coalescent point processes relies on laws of large numbers (see \cite{L2}).
The a.s. result for splitting trees relies on the theory of random characteristics (see \cite{CL1}) introduced in the seminal paper \cite{J} and further developed in \cite{JNa, JNb} and especially in \cite{Taib}.
\end{rem}
\begin{rem}
As in the last section, one could ask about the behaviour of large families, as the number $n$ of individuals grows. In contrast to the immigration case, here there are no families with abundances $O(n)$. Preliminary calculations \cite{CL2} show that there are two possible regimes, depending on the respective positions of the mutation rate $\theta$ and of the Malthusian parameter $\eta$ (see section on splitting trees). In the case when $\theta<\eta$ the abundance of the largest family is of order $O(n^{\beta})$, where $\beta=1-\theta/\eta$, otherwise it is of order $O(\log(n))$.
\end{rem}

As in the previous section, we have displayed results holding for a general lifespan measure $\pi$. On the other hand, here the quantities displayed in the theorem can only be computed in the case of critical birth--death processes, that is, when the death rate of individuals is constant, equal to their birth rate $\lbd$, so that $W(x) = 1+\lbd x$.
In that case,
$W_\theta'(x)= \lbd e^{-\theta x}$, and we can integrate the quantities in the theorem.

\begin{cor} In the case of a critical birth--death process with birth and death rate $\lbd$,
$$
\lim_{n\tendinfty} n^{-1}A_n(k) 	=	(\alpha^{-1}-1)\frac{\alpha^k}{k} \qquad \mbox{a.s.},
$$
where
$$
\alpha:=\frac{\lbd}{\lbd +\theta} .
$$
In addition,
$$
\lim_{n\tendinfty} n^{-1}A_n 	=	-(\alpha^{-1}-1)\ln (1-\alpha)\qquad \mbox{a.s.}
$$

\end{cor}


\begin{thebibliography}*

\bibitem{AN}
		{\textsc Athreya, K.B., Ney, P.E.}
		(1972)\\
		{\it Branching processes\/.}
		Springer-Verlag, New York.


	\bibitem{B}
		{\textsc Bertoin, J.}
		(1996)\\
		{\it Lévy processes.}
		Cambridge University Press, Cambridge.



\bibitem{CL1}
		{\textsc Champagnat, N., Lambert, A.}
		(2010)\\
		Splitting trees with neutral Poissonian mutations I: Small families. Submitted.

\bibitem{CL2}
		{\textsc Champagnat, N., Lambert, A.}
		(2010)\\
		Splitting trees with neutral Poissonian mutations II: Large families. In preparation.



	\bibitem{DT}	{\textsc Donnelly, P., Tavaré, S.}
	(1986)\\
	The ages of alleles and a coalescent. \emph{Adv. Appl. Probab.} {\bf 18} 1--19.



\bibitem{D} {\textsc Durrett, R.}
		(2008)\\
		{\it Probability Models for DNA Sequence Evolution.} Springer--Verlag, Berlin. 2nd revised ed.
		
		
		\bibitem{EAMcK}
	{\textsc Etienne, R.S., Alonso, D., McKane, A.J.}
	(2007)\\
	The zero-sum assumption in neutral biodiversity theory.
	 \emph{J. Theoret. Biol.} {\bf 248} 522--536. 
	


		
		\bibitem{Ew}
                {\textsc Ewens, W.J.}
                (1972)\\
                The sampling theory of selectively neutral alleles. \emph{Theoret. Popul. Biol.} {\bf 3} 87--112, and erratum, p.376.

		
				\bibitem{Ewens}
                {\textsc Ewens, W.J.}
                (2005)\\
                {\it Mathematical Population Genetics.} 2nd edition, Springer--Verlag, Berlin.


\bibitem{F}
	{\textsc Fisher, R.A.}
	(1943)\\
	A theoretical distribution for the apparent abundance of different species. \emph{J. Anim. Ecol.} {\bf 12} 54--58.

		

		
\bibitem{FCW}
	{\textsc Fisher, R.A., Corbet, S.A., Williams, C.B.}
	(1943)\\
	The relation between the number of species and the number of individuals in a random sample of an animal population. \emph{J. Anim. Ecol.} {\bf 12} 42--58.


	\bibitem{GK}
		{\textsc Geiger, J., Kersting, G.}
		(1997)\\
		Depth-first search of random trees, and Poisson point processes, in {\it Classical and modern branching processes } (Minneapolis, 1994) IMA Math. Appl. Vol. 84. Springer-Verlag, New York. 

\bibitem{HE}
	{\textsc Haegeman, B., Etienne, R.S.}
	(2008)\\
	Relaxing the zero-sum assumption in neutral biodiversity theory.
	 \emph{J. Theoret. Biol.} {\bf 252} 288--294. 

	\bibitem{H}
                {\textsc Hubbell, S.P.}
                (2001)\\
                {\it The Unified Neutral Theory of Biodiversity and Biogeography.} Princeton U. Press, NJ.



\bibitem{J}
		{\textsc Jagers, P.}
		(1974)\\
		Convergence of general branching processes and functionals thereof. \emph{J. Appl. Prob.} {\bf 11}  471--478.


\bibitem{JNa}
		{\textsc Jagers, P., Nerman, O.}
		(1984)\\
		The growth and composition of branching populations. \emph{Adv. Appl. Prob.} {\bf 16}  221--259.


\bibitem{JNb}
		{\textsc Jagers, P., Nerman, O.}
		(1984)\\
		Limit theorems for sums determined by branching processes and other exponentially growing processes. \emph{Stoch. Proc. Appl.} {\bf 17}  47--71.




\bibitem{KMcG}
                {\textsc Karlin, S., McGregor}
                (1967)\\
                The number of mutant forms maintained in a population. \emph{Proc. 5th Berkeley Symposium Math. Statist. Prob.} {\bf IV} 415--438.


\bibitem{Ken}
	{\textsc Kendall, D.G.}
	(1948)\\
	On some modes of population growth leading to R.A. Fisher's logarithmic series distribution. \emph{Biometrika} {\bf 35} 6--15.


\bibitem{KC}
	{\textsc Kimura, M., Crow, J.F.}
	(1964)\\
	The number of alleles that can be maintained in a finite population. \emph{Genetics} {\bf 49} 725--738.



\bibitem{L2}
		{\textsc Lambert, A.}
		(2009)\\
		The allelic partition for coalescent point processes. \emph{Markov Proc. Relat. Fields} \textbf{15} 359--386.


	\bibitem{L}
		{\textsc Lambert, A.}
		(2010)\\
		The contour of splitting trees is a Lévy process. \emph{Ann. Probab.} \textbf{38} 348–-395.



	\bibitem{McAW}
                {\textsc MacArthur, R.H., Wilson, E.O.}
                (1967)\\
                {\it The Theory of Island Biogeography.} Princeton U. Press, NJ.


\bibitem{Rannala96}
		{\textsc Rannala, B.}
		(1996)\\ 
		The sampling theory of neutral alleles in an island population of fluctuating size. \emph{Theoret. Popul. Biol.} \textbf{50} 91--104.
		

\bibitem{Rannala97}
		{\textsc Rannala, B.}
		(1997)\\
		Gene genealogy in a population of variable size. \emph{Heredity} \textbf{78} 417–-423.




\bibitem{Richard}
		{\textsc Richard, M.}
		(2010)\\
		Limit theorems for splitting trees with structured immigration and applications to biogeography. Submitted.


\bibitem{Taib}
{\textsc Taïb, Z.} (1992)\\
 {\it Branching processes and neutral evolution.} Lecture Notes in Biomathematics Vol. 93.
Springer-Verlag, Berlin.




\bibitem{W}
{\textsc Watterson, G.A.}
(1974)\\
Models for the logarithmic species abundance distributions. \emph{Theoret. Popul. Biol.} {\bf 6} 217--250.

\end{thebibliography}
\end{document}